# MEASUREMENT OF NORMAL CONTACT STIFFNESS ON FRACTAL ROUGH SURFACES

**Chongpu Zhai ***
School of Civil Engineering, The University of Sydney
Sydney, NSW, 2006, Australia. chongpu.zhai@gmail.com (Corresponding author)

**Sébastien Bevand**
Charles Delaunay Institute, LASMIS, University of Technology of Troyes (UTT)
UMR 6281, CNRS, 12 Rue Marie Curie, 10000 Troyes, France

**Yixiang Gan, Dorian Hanaor, Gwénaëlle Proust**
School of Civil Engineering, The University of Sydney
Sydney, NSW, 2006, Australia.

**Bruno Guelorget, Delphine Retraint**
Charles Delaunay Institute, LASMIS, University of Technology of Troyes (UTT)
UMR 6281, CNRS, 12 Rue Marie Curie, 10000 Troyes, France

## ABSTRACT

We investigate the effects of roughness and fractality on the normal contact stiffness of rough surfaces. Samples of isotropically roughened aluminium surfaces are considered. The roughness and fractal dimension were altered through blasting using different sized particles. Subsequently, surface mechanical attrition treatment (SMAT) was applied to the surfaces in order to modify the surface at the micro-scale. The surface topology was characterised by interferometry-based profilometry. The normal contact stiffness was measured through nano-indentation with a flat tip utilising the partial unloading method. We focus on establishing the relationships between surface stiffness and roughness, combined with the effects of fractal dimension. The experimental results, for a wide range of surfaces, showed that the measured contact stiffness depended very closely on surfaces' root mean squared (RMS) slope and their fractal dimension, with correlation coefficients of around 90%, whilst a relatively weak correlation coefficient of 57% was found between the contact stiffness and RMS roughness.

## INTRODUCTION

In contact mechanics, surface morphology and features play a considerable role in determining how solids will interact with one another in processes including friction, wear, lubrication, thermal and electrical conductance (Bowden et al. 1986; Johnson 1987). Rough surfaces can be characterised using a variety of descriptors including surface roughness, e.g., root mean squared roughness, skewness, and fractal dimension. Surface fractal dimension is regarded as an effective roughness parameter since it represents three-dimensional surface features over a significant range of length scales (Go et al. 2006). In the past years, investigations have been done extensively regarding the normal interfacial stiffness of random rough surfaces with varying fractal dimensions. Numerical simulations have been carried out with the help of molecular dynamics (Akarapu et al. 2011), boundary element methods (e.g., Pohrt and Popov 2012, 2013; Campana et al. 2011), etc. Studies have focused on interpreting parameters of contact area and contact stiffness under different loading conditions and surface characteristics (Popov 2010). Several related experiments have also been conducted on rough surfaces to obtain interfacial behaviour (e.g. Buzio et al. 2003; Lorenz et al. 2009; Hanaor et al. 2013). In spite of recent technological advancements in nano-indentation and surface morphology characterisation, which facilitate the mechanical and morphological analysis of surfaces at micrometer and nanometer scales, the structure-dependence of contact behaviour of surfaces with random multiscale features remains largely unknown, and particularly the existing experimental results are limited. In this work, we present an experimental study to determine correlations between interfacial contact stiffness and surface morphology.

## METHOD

### Sample Preparation

Thick round disks made of aluminium alloy 2011 were selected as testing samples due to the material's chemical stability and deformability. The surfaces of these samples were firstly treated by standard sand blasting procedures. The sand particles used to blast the sample surfaces were garnet and glass beads, the sizes of which are 0.2 mm and 0.3 mm, respectively. SMAT treatment utilising the vibration and shock of micro-scale particles was further employed on certain samples in order to further modify surface features.

The surface morphology of the testing samples is of fundamental importance in governing physical properties and interfacial phenomena. We described the sample surfaces through descriptors of RMS roughness as well as RMS slope, which are two widely used parameters for rough surfaces (Hanaor et al. 2013). For the purpose of quantifying the surface roughness, the aluminium surfaces were scanned using an optical surface profilometer (NanoMap 1000WLI) to obtain three-dimensional topographies of the samples. Then, the mean roughness parameters were determined from the digitised surface data across multiple scans of $1\times1$ mm$^2$ for each individual sample.

As RMS roughness is influenced primarily by highest level features while RMS slope values are influenced primarily by lowest scale features, these parameters are rather insufficient to give an entire description of the surface over multiple length scales. The concept of fractal dimension provides a useful method for characterising rough surfaces in terms of cross-scales analysis. A number of existing algorithms have been developed to determine surface fractal dimension from digitised surface data. Here we adopted a triangulation method to calculate the fractal dimension of a given surface scan. The method works as shown in Fig. 1. A grid of unit dimension $L$ is placed to mesh the surface into a number of triangles. For example, when $L = X/4$, with $X$ being the total scan length, the surface is covered by 32 triangles of different areas inclined at various angles with respect to the projected plane. These triangles have an equal projected triangle size, although their real areas are different and larger than the projected area $A_P$. The areas of all triangles are calculated and summed to obtain an approximation of the surface area $A_S(L)$ depending on $L$. The grid size is then decreased by a successive factor of 2, and the previously mentioned process continues until $L$ corresponds to the distance between two adjacent pixel points, i.e., the highest resolution of the scan. The fractal dimension $D_f$ can be calculated (Douketis et al. 1995) as follows:

$$D_f = -\frac{d\log(A_S / A_P)}{d\log(L)} + 2. \tag{1}$$

As evident from Fig. 1, the sample surfaces exhibit self-affinity only over a certain range of scales, following the tendency of natural surfaces to generally have complex structures with self-affinity (statistical self-similarity) after anisotropic scaling (Zahn et al. 1999). We used the anisotropic scaling to normalise three-dimensional scans to calculate $D_f$ for characterising the sample surfaces.

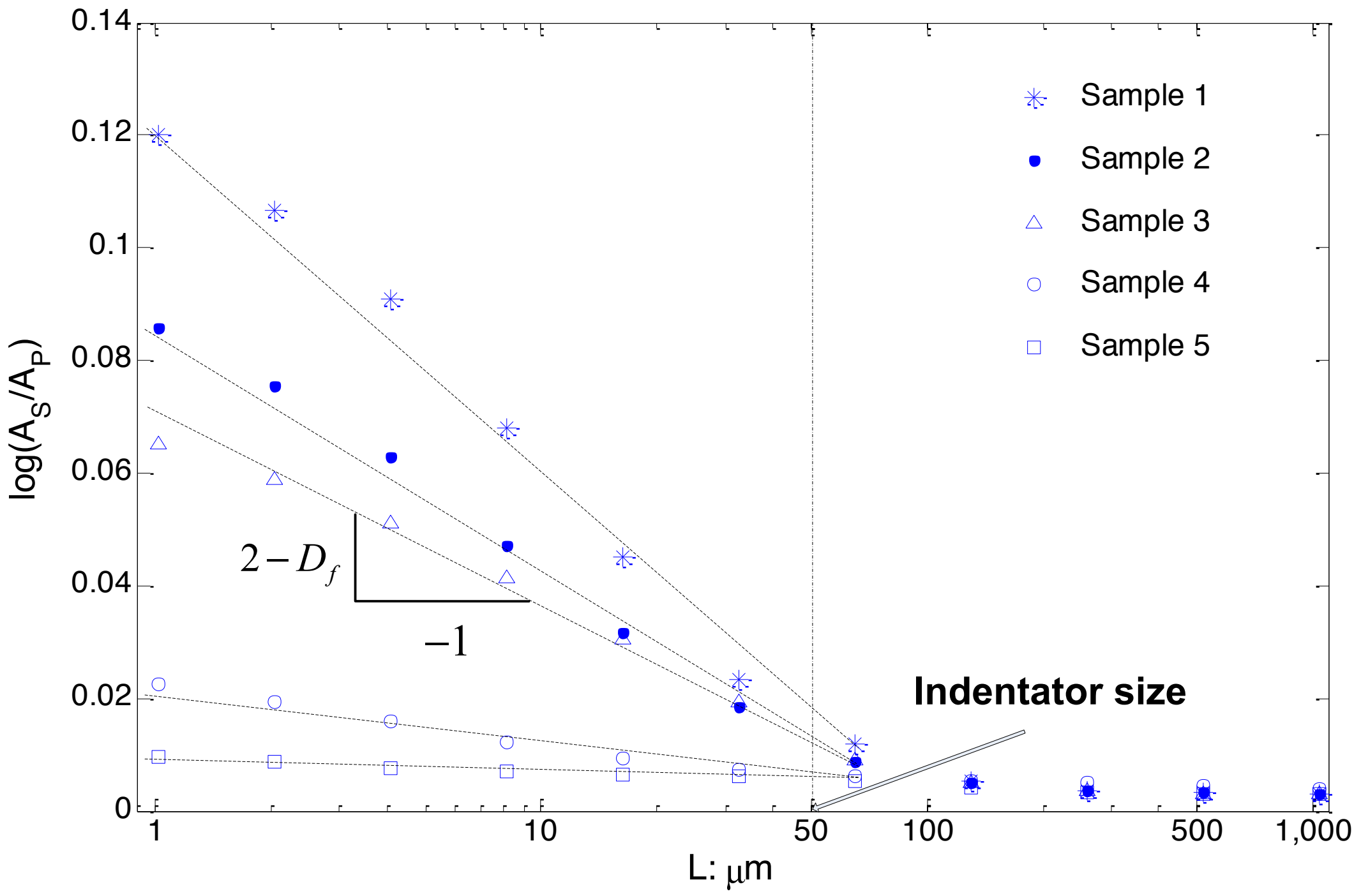

Figure 1. Schematics of the calculation of $D_f$ from different sample surfaces. The size of flat tip used in nano-indentation is also shown. Sample details can be found in Table 1.

As shown in Table 1, five typical sample surfaces with distinct roughness parameters were prepared and characterised prior to nano-indentation tests. For each sample, the mean RMS roughness, RMS slope, and fractal dimension are reported with standard deviations as indicated in Table 1. It is found that the RMS slope and fractal dimension reveal similar changing trends, i.e., with smaller particles used to modify the surfaces, larger RMS slope and fractal dimension values are achieved. As for the RMS roughness, results show that the parameter has no clear correlation with varying particle sizes.

Table 1: Surface characterization with different treatments.

| Surfaces | RMS roughness/μm | RMS slope | Fractal dimension |
|---|---|---|---|
| Sample 1: Surface blasted with garnet (particle size: 0.2 mm) | 4.075 ± 0.404 | 0.24 ± 0.011 | 2.69 ± 0.064 |
| Sample 2: Surface blasted with glass beads (particle size: 0.3 mm) | 4.716 ± 0.372 | 0.22 ± 0.08 | 2.61 ± 0.037 |
| Sample 3: Surface blasted with garnet + SMAT (particle size: 2 mm) | 4.422 ± 0.597 | 0.19 ± 0.009 | 2.55 ± 0.042 |
| Sample 4: Surface blasted with glass beads + SMAT (particle size: 2 mm) | 3.554 ± 0.958 | 0.14 ± 0.007 | 2.30 ± 0.051 |
| Sample 5: Polished surface with SMAT treatment (particle size: 2 mm) | 3.641 ± 0.776 | 0.07 ± 0.006 | 2.26 ± 0.073 |

**Nano-indentation**

Surface contact stiffness of aluminium samples with different surface morphology was assessed through nano-indentation tests with a 54 μm-diameter flat tip (SYNTON-MDP FLT-D050), as shown in Fig. 2. The tip size has a length scale comparable with the surface correlation length, shown in Fig. 1. During the tests, the flat tip penetrated into random rough samples. We chose a flat tip so that the apparent contact area under the tip does not change with respect to the indentation depth, unlike spherical and Berkovich tips. All the values of stiffness were gained by averaging over 10 indentation tests at different positions within each sample. Unloading processes at predetermined loading levels were applied to eliminate the influence of plastic deformation while measuring contact stiffness. For each indentation test, 10 partial unloading procedures were employed with 10% of the current loading level.

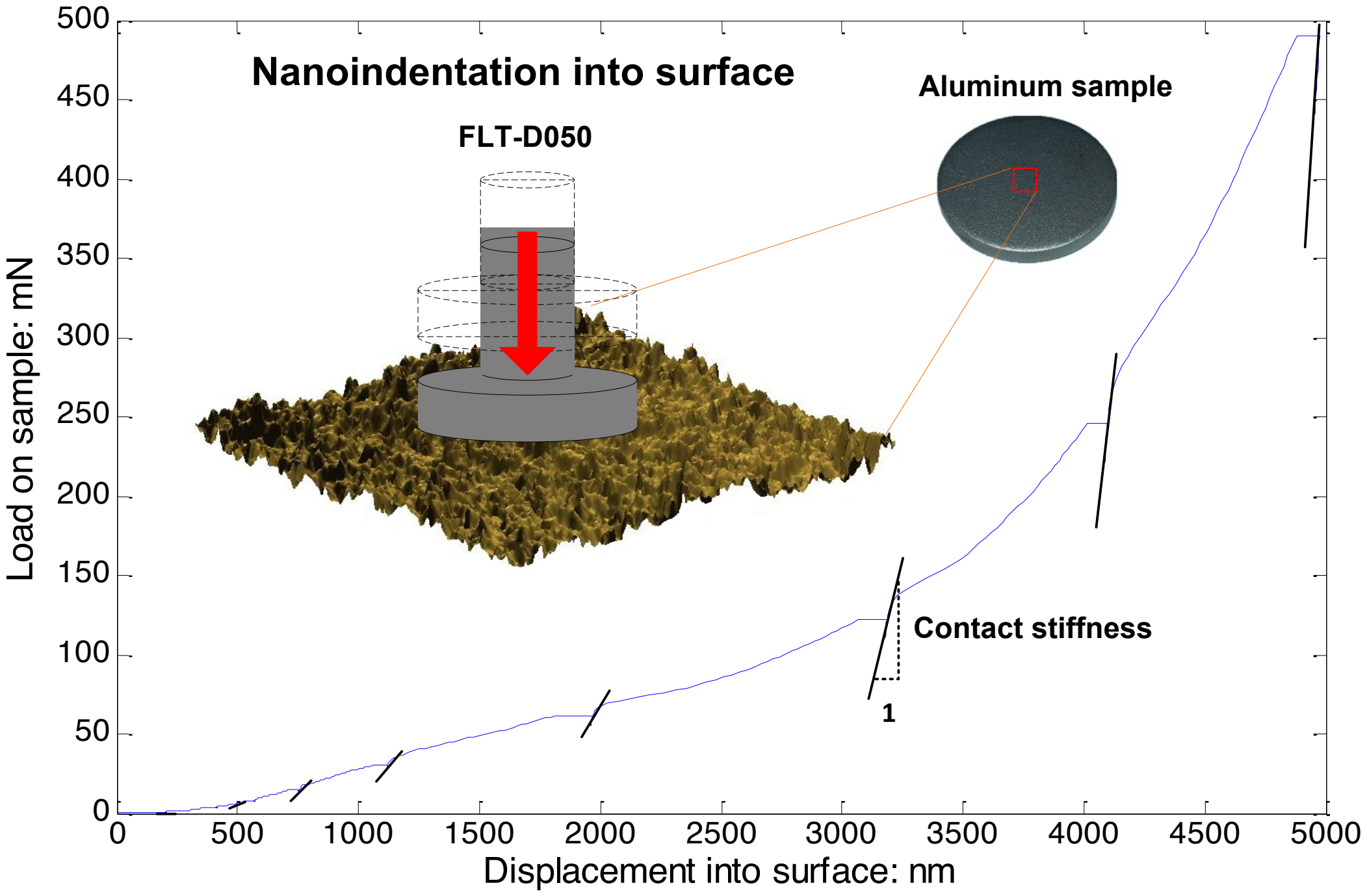

Figure 2. Typical loading curve of nano-indentation tests on the sample surfaces. Partial unloading paths were applied to extract surface contact stiffness under different loading levels.

## RESULTS AND DISCUSSIONS

### Experimental Data

Fig. 3(a) shows the contact stiffness of five selected surfaces for different values of stress $\sigma$, which is calculated as loading force divided by projected area of the tip. For all tested samples, the results agreed well with the existing theory that the contact stiffness increases with the increment of loading force. It is also found that a high RMS slope and fractal dimension results in a low contact stiffness value. To further clarify the specific differences of contact stiffness between distinct surfaces, we converted the raw data to non-dimensional values by dividing the stiffness and stress values ($k_{min}$ and $\sigma_{min}$, respectively) by their values at the first unloading stage for each surface. It can be seen in the low stress region (i.e., $s_{min} < 100$ MPa), of Fig. 3(b) that the logarithm of dimensionless stiffness is proportional to the logarithm of the dimensionless normal stress. That is to say, the stiffness shows a power-law relationship with the normal force, which can be described as

$$k : F_N^{\alpha}, \tag{2}$$

where $F_N$ is the normal force acting on the surface and $\alpha$ is the exponent of the power function. The stiffness exponent $\alpha$ ranges from 0.6407 to 0.7873, changing as the fractal dimension $D_f$ changes from 2.26 to 2.69 and the RMS slope from 0.07 to 0.24. As a comparison, the typical value of for Hertzian contact is 1/3.

The power-law relationship found in these experimental results is consistent with previous theoretical predictions (Pohrt et al. 2012) on a quantitative basis. The results obtained here also correspond well with results from simulations (Geike et al. 2007) and experimental measurements carried out through AFM on thin films (Buzio et al. 2003).

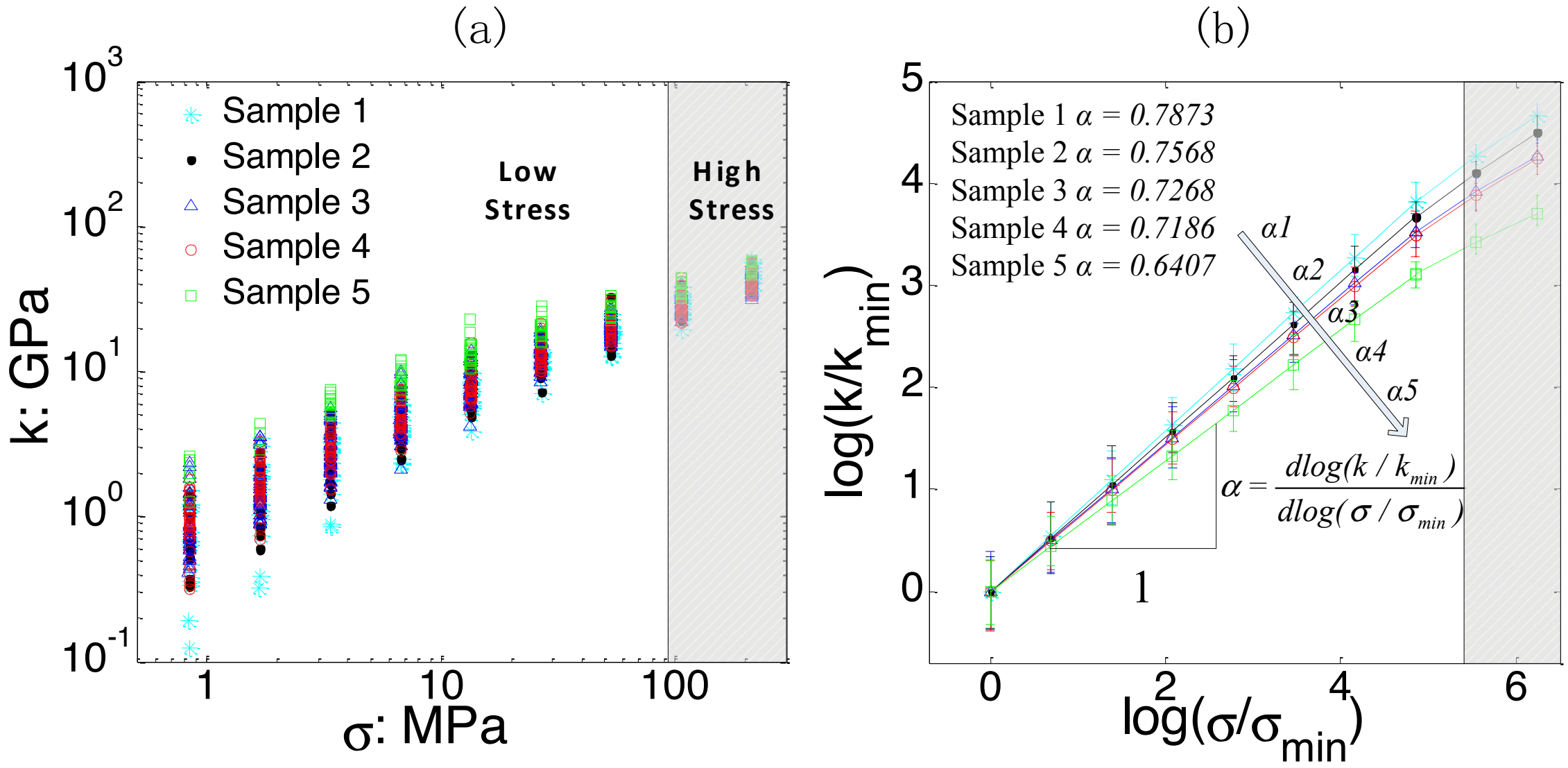

Figure 3. Contact stiffness of five tested samples: (a) Raw data from multiple indentation tests at different loading levels, and the stress is calculated by $F/A$ where $F$ is the normal force and $A$ denotes the projected area of the flat tip; (b) Curve fitting for normalised stiffness and stress for each individual sample, with $k_{min}$ and $\sigma_{min}$ being stiffness and stress at the first unloading stage, respectively.

We emphasise here that the above observations are valid primarily for the low stress region. As the loading force increases towards the high stress region, the exponent $\alpha$ exhibits a slight downward trend. The trend is more profound as the normal force approaches the compressive strength of the aluminium alloy. The non-uniform pressure distributions within the real contact areas remain non-linear and complex in the plasticity-dominant regions. The impact of plastic deformation of individual asperities can be a possible explanation for this trend.

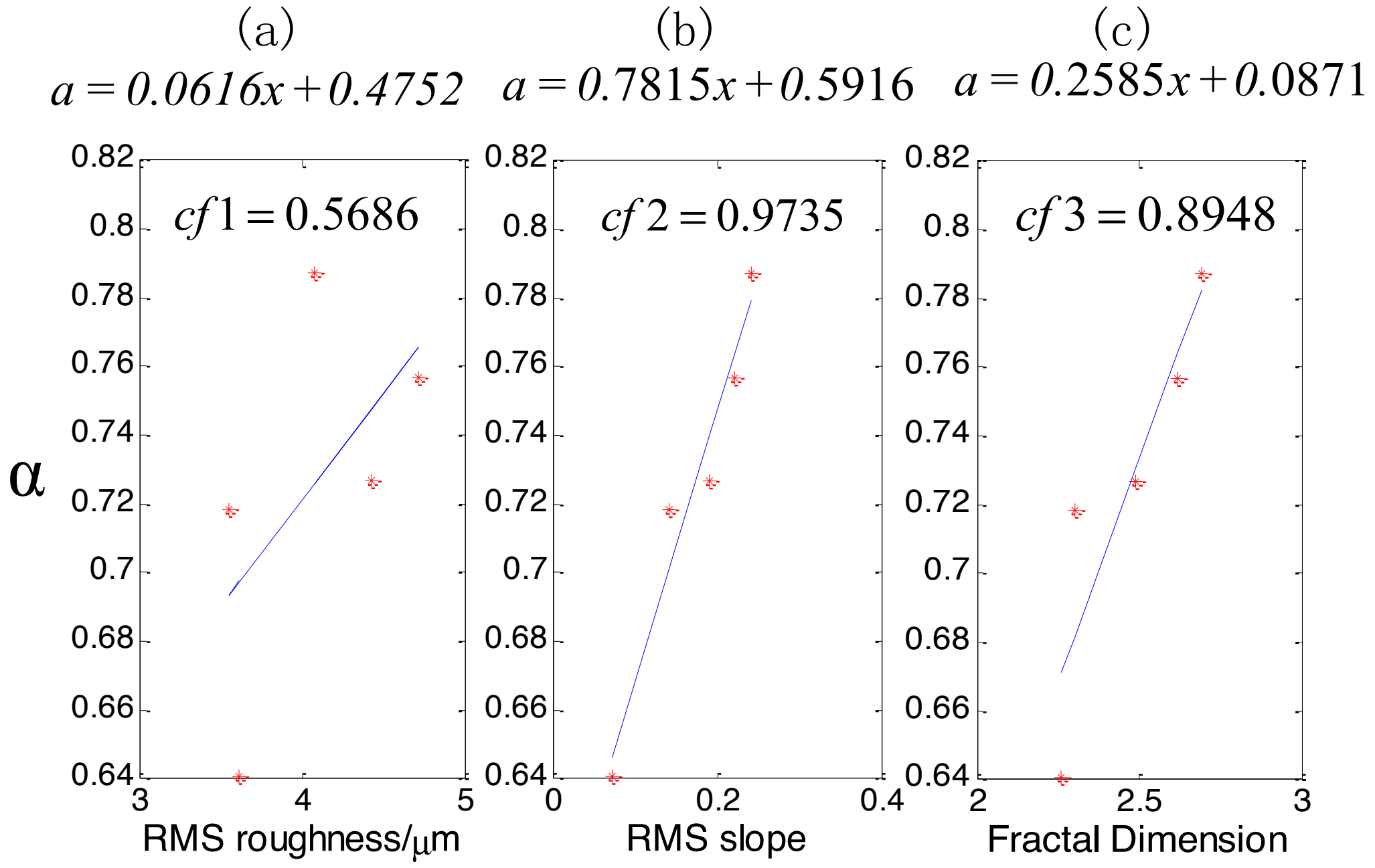

Figure 4. Correlations between the stiffness variation exponent $\alpha$ and roughness parameters: (a) RMS roughness; (b) RMS slope; (c) Fractal dimension.

**Correlation Analysis**

Fig. 4 illustrates the significance of RMS roughness, RMS slope and fractal dimension in governing contact stiffness for rough surfaces. Specifically, the stiffness variation exponent $\alpha$ across different loading forces illustrates a relatively weak correlation to RMS surface roughness. While in contrast, the exponent correlates more closely with the RMS slope and the fractal dimension with the corresponding correlation coefficients around 90%.

## CONCLUSIONS

We experimentally demonstrate the effects of surface roughness on the normal contact stiffness of various rough surfaces. The results in our experiments show that the contact stiffness follows a power-law function with respect to the normal force. The exponents of stiffness variation over loading show strong dependence on the RMS slope and fractal dimension. Some further research can be carried out with various tip sizes to explore cross-scale contact properties.

## ACKNOWLEDGMENTS

Financial support for this research from the Australian Research Council through grants DE130101639 and Civil Engineering Research Development Scheme (CERDS) in School of Civil Engineering at The University of Sydney is greatly appreciated.